\documentstyle[10pt,aaspp4]{article}

\newcommand{\bq}{\begin{equation}}
\newcommand{\eq}{\end{equation}}

\begin{document}
\def\refitem{\par\parskip 0pt\noindent\hangindent 20pt}
\normalsize

\title{$BVRI$ Light Curves for 22 Type Ia Supernovae}

\bigskip

{\it Accepted to the Astronomical Journal}

\vspace*{0.3cm}

Adam G. Riess\footnote{Department of Astronomy, University of California, 
Berkeley, CA 94720-3411},
Robert P. Kirshner\footnote{Harvard-Smithsonian Center for Astrophysics, 60
Garden St., Cambridge, MA 02138},Brian P. Schmidt\footnote{Mount Stromlo and Siding Spring Observatories,
Private Bag, Weston Creek P.O. 2611,  Australia},Saurabh Jha$^2$,
Peter Challis$^2$, Peter
M. Garnavich$^2$, Ann A. Esin$^2$, Chris Carpenter$^2$,
Randy Grashius\footnote{University of New Mexico, Capilla Peak
Observatory Abuquerque, NM 87131}, Rudolph E. Schild$^2$, Perry L. Berlind\footnote{Fred
Lawrence Whipple Observatory, Amado, AZ 85645}, John P. Huchra$^2$,
Charles F. Prosser\footnote{NOAO, Tucson, AZ 85726}, Emilio
E. Falco$^2$, Priscilla J. Benson\footnote{Whitin Observatory,
Wellesley College, MA 02481-8203}, Cesar Briceno$^2$, Warren
R. Brown$^2$, Nelson Caldwell$^5$,  Ian P Dell'Antonio\footnote{Bell
Lab, Murray Hill, NJ 07974}, Alexei V. Filippenko$^1$, Alyssa
A. Goodman$^2$, Norman A. Grogin$^2$, Ted Groner$^5$, John P. Hughes\footnote{Rutgers University, Dept. of Physics \& Astronomy, New Brunswick, NJ 08855}, Paul
J. Green$^2$, Rolf A. Jansen$^2$, Jan T. Kleyna$^2$, Jane X. Luu$^2$, Lucas M. Macri$^2$,
Brian A. McLeod$^2$, Kim K. McLeod$^7$,Brian R. McNamara$^2$,
 Brian McLean\footnote{Space Telescope Science Institute, Baltimore, MD
21218}, Alejandra A. E. Milone\footnote{Multiple Mirror Telescope Observatory,
c/o Whipple Observatory,P.O. Box 97, Amado AZ 85645-0097},Joseph J Mohr\footnote{Department of Astronomy and Astrophysics,
University of Chicago, Chicago, IL  60637}, Dan Moraru$^2$, Chien Peng$^{1}$\footnote{Steward Observatory, University of Arizona, Tucson, AZ 85721}, 
Jim Peters$^5$, Andrea
H. Prestwich$^2$, Krzysztof Z. Stanek$^2$, Ping Zhao$^2$
\setlength{\baselineskip}{22pt}
\setlength{\parskip}{0.5ex}

\begin{abstract}
    
We present 1210 Johnson/Cousins B,V,R, and I photometric
observations of 22 recent type Ia supernovae (SNe Ia): SN 1993ac, SN 1993ae, SN 1994M, SN 1994S, SN 1994T, SN 1994Q, SN 1994ae, SN 1995D, SN 1995E, SN 1995al, SN 1995ac, SN 1995ak, SN 1995bd, SN 1996C, SN 1996X, SN 1996Z, SN 1996ab, SN 1996ai, SN 1996bk, SN 1996bl, SN 1996bo, and SN 1996bv.  Most of the photometry was obtained at the Fred Lawrence Whipple Observatory (FLWO) of the Harvard-Smithsonian Center for Astrophysics in a cooperative observing plan aimed at improving the data base for SN Ia.  The redshifts of the sample range from $cz$=1200 to 37000 km s$^{-1}$ with a mean of $cz$=7000 km s$^{-1}$. 

\end{abstract}

\vfill
\eject

\section{Introduction}

Recent evidence suggests that type Ia supernovae (SNe Ia) can be used as exceedingly precise long-range distance indicators (Riess, Press, \& Kirshner 1995a,b, 1996a,b; Hamuy et al. 1995,1996a,b; Maza et al. 1994; Phillips 1993; Tammann \& Sandage 1995).  With peak luminosities a million times greater than Cepheid variables and individual distance accuracy approaching 5\%, they provide cosmology with a tool of great leverage. 
 
 The uses for these extragalactic beacons are numerous.  As test
 particles in the nearby Hubble flow, they have been used to measure
 the current expansion rate of the Universe (Sandage et al. 1992, 1994,
 1996; Sandage \& Tammann 1993; Schaefer 1994, 1995a,b, 1996; Branch
 \& Tammann 1992; Tammann \& Leibundgut 1990; Arnett, Branch, \&
 Wheeler 1985; Cadonau, Sandage, \& Tammann 1985; Hamuy et al. 1995;
 1996a,b; Riess, Press, \& Kirshner 1995a, 1996a; Riess et al. 1998a;
 Tripp 1998; Branch 1998 and references within).
  Combined with their
 positions on the sky, SNe Ia have been used to reveal the peculiar
 velocities of distant galaxies as well as the bulk flow of our own
 local neighborhood (Tammann \& Leibundgut 1990; Miller \& Branch
 1992; Jerjen \& Tammann 1993; Riess, Press, \& Kirshner 1995b;
 Watkins \& Feldman 1995; Riess et al. 1997b; Zehavi et al. 1998;
 Tammann 1998).
  SN Ia evolution is the only well
 understood time-variable process which can be used to mark the
 passage of time at high redshift.  As cosmological clocks, SNe Ia have
 been used to examine the nature of the redshift using the time
 dilation test (Rust 1974; Leibundgut 1990; Goldhaber et al. 1997;
 Leibundgut et al. 1996; Riess et al. 1997a).  SNe Ia have been
 employed as probes of extragalactic dust (Della Valle \& Panagia
 1993; Riess, Press, \& Kirshner 1996b), and their contribution to
 galactic chemical enrichment by their production of iron peak
 elements has been explored by measuring their rates of occurrence
 (Timmes 1991; Cappellaro et al. 1993a,b, 1997; Turatto et al. 1994; van den Bergh
 \& McClure 1994; Pain et al. 1997; Madau, Della Valle,
 \& Panagia 1998).
 Recently, vigorous programs have embarked on searches for SNe Ia at
 high redshifts (0.2 $\leq$ $z$
 $\leq$ 1.0) with the intent of measuring the expansion history of the
 Universe (Perlmutter et al. 1995, 1997; Schmidt et
 al. 1998).  Early results imply that there is not enough 
 gravitating matter to close the Universe (Garnavich et al.
 1998a; Perlmutter et al. 1998) and that currently the expansion is 
 accelerating (Riess et al. 1998b; Perlmutter 1999).  Supernova observations, when
 combined with measurements of Cosmic Microwave Background
 anisotropies, may prove useful to determine the cosmic equation of
 state and the global geometry of the Universe (Garnavich et al. 1998b;
 White 1998).

   These applications require well-observed SN Ia light curves
   with reliable photometry.  Further, the cosmological applications
   rely on comparisons to nearby SNe Ia which delineate today's Hubble flow ($0.01 < z <
   0.1$). 

 Despite the importance of precisely observed SN Ia light curves, most published SN
   Ia photometry, before 1980, consisted of infrequently
   sampled photographic light
   curves for SNe within $cz < 2000$ km s$^{-1}$ with a wide assortment of filters and emulsions (van den
   Bergh 1994; Cadonau \& Leibundgut 1990; Barbon, Cappellaro, \&
   Turatto 1989).
   Much of this data is plagued by systematic
   errors from non-linearities in detector sensitivity, uncertain
   transformations to modern filter conventions, and difficulties with
   background light subtraction.  
A crude estimate of these
   photometry errors comes from comparing the dispersion in Sandage \& Tammann's (1993)
   Hubble diagrams constructed with old photographic
   light curves (0.65 mag) with those based on more modern $B$-band SN Ia observations
   (0.38 mag; Hamuy et al. 1996a).  The result suggests that typical
   errors from the pre-1980 photographic SN Ia photometry could be as
   high as $\sim$ 0.5 magnitudes.  

    Obtaining well sampled light curves with high-precision
 photometry ($\sigma
    \leq $ 0.03 mag) is challenging.  Collecting
    observations of SNe Ia in each of four filters every few days for
    $\sim$ 100 days is a task that is not well suited to the short blocks of
    time allocated at most modern observatories.  Variable weather
    poses another challenge to obtaining well-sampled SN Ia light
    curves. It is also important to maintain a telescope-detector 
   setup throughout the observations which well matches standard passband
  conventions (Johnson \& Harris 1954; Bessell 1990) since the non-stellar
 spectrum of an SN Ia can make linear color corrections inexact.
  The most challenging obstacle to producing a high-quality SN Ia light curve
 is to account
 correctly for the background light from the host galaxy {\it at the
 position of the supernova}.  

Recent evidence has shown that SNe Ia are not perfectly homogeneous
     in luminosity or color and that the intrinsic luminosity and
     color is intimately related to the {\it shape} of the observed
     light curves (Phillips 1993; Riess, Press, \& Kirshner 1995a, 1996a;
     Riess et al. 1998b; Hamuy et al. 1995, 1996a).
     Incorrectly subtracting the background light from a set of
     supernova observations can have disastrous effects on the light
     curve shape.  Boisseau \&
     Wheeler (1991) have investigated the effect of background galaxy
     contamination on the inferred absolute magnitude and light-curve
     speed of SNe Ia.  They find that oversubtraction or
     undersubtraction of a constant flux source leads to an observed
     correlation between SN Ia light curve speed and inferred
     luminosity.  If the value of this correlation were the same as the
     intrinsic correlation, then using the light-curve shape to
     correct the luminosity could equally well account for either intrinsic
     luminosity variation or galaxy contamination.  Unfortunately,
     they are not the same, so it is crucial to account correctly for
     the background light to determine the true light-curve shape.  SN
     Ia light curves obtained by the Cal\'{a}n/Tololo survey show that
     with high quality photometry, it is possible to measure distances
     with SN Ia light curves to a precision approaching $\sim$5\% (Hamuy et al.
     1995, 1996a; Riess, Press \& Kirshner 1995a, 1996a).  

Since the widespread use of modern
   CCD detectors coupled with commercially available Johnson/Cousins
   passbands began in $\sim$ 1980, there have been nearly 250 SNe Ia
   reported (van den Bergh 1994; Barbon, Cappellaro, \& Turatto
   1989).  Regrettably, light curves have been collected,
   reduced, and published for fewer than 50 of these objects (Cadonau
   \& Leibundgut  1990; Hamuy et al. 1993; Hamuy et al. 1994; 1996b;
   Sadakane et al. 1996; Phillips 1993 and references within).  Of
   these, less than half include $I$ and $R$ light curves which, combined
   with shorter wavelength light curves, can be used to determine the reddening
   due to dust (Ford et al. 1993; Hamuy et al. 1996b).  

   Recently, progress has been made toward building a reliable sample
   of SN Ia light curves in the Hubble flow.  The largest
   contribution to date has been made by the Cal\'{a}n/Tololo
   Supernova Survey, a program begun in 1990 by astronomers at Cerro
   Tololo Inter-American Observatory (CTIO) and the Cerro Cal\'{a}n
   Observatory of the University of Chile (Hamuy et al. 1993).  This
   photographic search with follow-up $B$,$V$,$I$ CCD photometry
   netted 27 SNe Ia with ($0.01 < z < 0.1$).  Other programs
   which show great promise include the Beijing Astronomical
   Observatory search (IAUC 6379), the Mount Stromlo Abell Cluster
   Supernova Search (Reiss et al. 1998), and the Lick Observatory
   Supernova Search (IAUC 6627), all of which have made repeated
   discoveries
   and in the future are expected to
   contribute to the growing sample of SN Ia light curves.

   Beginning in 1993, astronomers at the Center for Astrophysics
  (CfA) began a concerted and organized effort to collect
  Johnson/Cousins $BVRI$ photometry of type Ia supernovae.
  Many of these SNe Ia were discovered serendipitously by amateurs or 
  by professionals scanning images collected for another
  purpose.  This work presents the light curves of 22 of SNe Ia in the
  Hubble flow observed between 1993 and 1996.  In \S 2 we give details
  of our observational setup and reduction
  procedure.  We present $BVRI$ photometry for 22 SNe Ia in \S 3.  In
  \S 4 we discuss the characteristics of this sample.

\section{Observations \& Reductions}
  Most of the photometric data presented in this paper were collected
   at the 1.2 m telescope at the Fred Lawrence Whipple Observatory (FLWO).
   The 1.2 m is an $f/8$ Ritchey-Chretien reflector and was outfitted with a thick
   front-illuminated Loral CCD between 1993 and July 1995, and a
   thinned, back-side illuminated Loral CCD from August 1995 through 1996.
  The surfaces of both CCDs were coated with
   a laser dye which improves the blue sensitivity.  The pixels are 15 microns
   square, corresponding to 0.31 arcseconds at the focal plane of the 1.2 m telescope.  The filters are constructed from Schott glass components recommended by Bessell (1990) for a coated CCD.

   The $B,V,R,$ and $I$ CCD transmission functions for the FLWO 1.2 m telescope are shown in
   Figure 1 (Andy Szentgyorgyi, private communication).   The CCD's ability to detect light over a range of wavelengths makes it difficult to emulate the sharp blue-side or red-side cut-offs of photomultiplier transmission functions.  The $B$ and $V$ CCD transmissions closely match the photomultiplier passband conventions of
   Johnson \& Harris (1954) and most recently Bessell (1990).  The $R$-band CCD transmission is similar to Cousins (1980, 1981) and Bessell (1990). The $I$-band CCD transmission extends to substantially longer wavelengths than
   the Cousins (1980, 1981) and Bessell (1990) convention for the $I$ photomultiplier passband.  
    The FLWO CCD transmissions are very similar to CCD transmission functions obtained by Bessell with the filters he prescribed (Bessell 1990).  We use linear color corrections to account for differences between our CCD transmission functions and the Johnson/Cousins photomultiplier passbands.   Still, broad emission and absorption features in the spectral energy distribution of SN Ia
   can cause variations among light curves observed with slightly different CCD transmission functions.
   The difference between broad-band SN Ia photometry obtained at FLWO and
   CTIO has been determined in detail by Smith et al. (1998) by
   comparing phototmetry of SN 1994D obtained at the two sites.  These
   differences (with uncertainties in parenthesis), as shown in Table 1, are small, but not entirely absent.  Agreement between FLWO and CTIO is best in the $V$ passband.

        A small number of the observations reported here were conducted at the 0.62 m
   telescope at the Capilla Peak Observatory (CPO).  The 0.62 m
   is a $f/15.2$ Boller \& Chivens Cassegrain telescope matched with
   a RCA model SID501EX back-illuminated and thinned CCD.
   The CCD pixels are 30 microns square, corresponding to 0.67 arcseconds
   per pixel at prime focus.  The CPO CCD transmission functions are
   shown by Beckert \& Newberry (1989) and are a good match to the
   CCD transmissions at FLWO.  The mean difference in SN Ia photometry
   obtained at FLWO and CPO is shown in Table 1 using SN 1995D.  The
   difference is very small in $V$ and somewhat larger in $B,R,$ and $I$.
   Approximately 10 of the more than 1200 photometric observations
   were collected at other observatories including the
   Michigan-Dartmouth-MIT Observatory, the McDonald Observatory, 
the McGraw Hill Observatory, the Lick Observatory, and CTIO as noted in the photometry tables.

A well-sampled SN Ia light curve requires monitoring every day or two near 
maximum light and every few days as it changes more slowly a fortnight
after maximum.  Weather and moonlight often intercede even when an observatory is well-organized and well-instrumented to gather these observations.  Fortunately, 
photometric weather is not required since we can use stars in the SN field as local calibrators to monitor the changing brightness of the SN Ia through differential photometry.
 Optimal comparison stars are the brightest stars in the
   field which do not saturate the CCD electron wells during the long SN
   Ia exposures at late times when the supernova has dimmed.
   Comparison stars obviate the need to make first-order airmass
   corrections and allow the SN Ia brightness to be measured
   in any weather conditions for which the supernova is visible.
 
      On nights which are photometric, we performed all-sky photometry
      from which we constructed a transformation from
   our detector measurements to the standard photometric
   conventions.  Following Hardie (1962) and Harris et al. (1981), we used
   transformation equations which gave the apparent magnitude
   as a linear combination of the instrumental magnitude,
   observed airmass, and color.  Using all-sky standard stars from Landolt
   (1992) we then solved for the linear coefficients of the
   transformation equations.  The typical $rms$ scatter of our
   transformations was 0.02 mag, with no observed
   correlation between residuals and color, airmass, or instrumental magnitude.
  The mean color terms for the FLWO 1.2 m were 0.04, -0.03, -0.08, and 0.06 mag in $B,V,R,$ and $I$ per mag of $B-V$, $B-V$, $V-R$, and $V-I$, respectively.
   These transformations were employed to calibrate the apparent magnitudes of the comparison stars which were observed on the same night
   as the Landolt standards.  

 The comparison stars used for each SN Ia
   field are marked in Figure 2 and their $B,V,R,$ and $I$ apparent
   magnitudes are given in Table 2.  For two fields (as listed in
   Table 2), only one ``primary'' comparison star was consistently
   visible in the field of view.

   For each night a SN Ia was observed, we measured the brightness
of the supernova relative to one or more comparison
stars in the field.  Extreme care was exercised to subtract properly
the background light at the location of the
supernova.  For SNe Ia far from the galaxy, or on a smooth and uniform region of
the galaxy, reliable background subtraction was easily accomplished.  In such cases,
we generally measured the background light from the median sky value contained
in an annulus of width $\sim$ 6$^{\prime\prime}$ at an inner distance of $\sim$ 8$^{\prime\prime}$ from the
center of the supernova.  These separations were increased as needed
for data from nights with poor seeing.
More challenging were SNe Ia located on a luminous and mottled galaxy background.  In this case, unless images of the galaxy existed prior
to explosion, the only reliable way to proceed was to
wait for the SN Ia to fade away to obtain an image of the galaxy without the supernova.
Then by carefully matching the alignment, intensity, and point-spread
functions of the images with and without the SN Ia present, we
subtracted the two images to obtain an image of the supernova with zero
background (Schmidt et al. 1998).  This is the same method used for
the photometry of high-redshift SNe Ia (Riess et al. 1998b).   

To measure the brightness of the supernova relative to the comparison
stars in an uncrowded field we used the method of aperture photometry.
We added the light contained in a series of apertures of increasing
radius around the star and supernova and found the difference in
magnitude between the SN Ia and the comparison star which was
independent of aperture radius.  The estimated background was varied
until a difference in magnitude was found which was independent of aperture radius. This procedure is refined for crowded fields or faint SNe Ia where we have fit point spread functions (PSFs) to the
SN Ia and comparison stars to determine their relative brightness.
Experience has shown that when either technique is suitable, the
aperture method and the PSF method give consistent results within $0.01$ mag.  The
particular method used to derived the photometric measurements for each SN is listed in Table 3.

\section{SN Ia Data}

     $B,V,R,$ and $I$ band photometry for 22 SN Ia light curves is
     given in Tables 6 through 27 and plotted in Figure 3.
 For each observation
     we include an estimate of the 1 $\sigma$ error which was
     determined from Poisson statistics, image quality, and
     uncertainty in the calibration of the comparison stars.
     In most cases the dominant source of uncertainty is the
     comparison star calibration.  
    In Table 3 we give details relevant to the SN Ia
     observations including heliocentric
     redshift (column 2), the peak of the $B$ light curve (column 2), the peak of the $V$ light curve (column 3), the decline in $B$ 15 days after maximum, $\Delta_{m15}(B)$ (column 4), time of the first observation relative to $B$
     maximum (column 5) as determined by the multicolor light curve shape (MLCS) fit, right ascension (column 6), declination
     (column 7), and the photometric reduction technique (column 8)
     used to measure the SN Ia's brightness.  The redshifts for SNe
     1994S, 1994ae, 1995D, 1995E, 1995al, 1996Z, 1996ai, 1996bo, 1996bk, and 1996bv are from Huchtmeier
     \& Richter (1989); SN 1993ae is from Chincarini \& Rood (1977);
     SN 1995ak is from IAUC 6254; SN 1996bo is from IAUC 6492; and SN
     1996X is from the RC3 catalogue (de Vaucouleurs et al 1991).  The rest were determined from our spectra of the host galaxies. 

 The peaks of the $B$ and $V$ light curves and values of $\Delta_{m15}(B)$ were determined from a light curve fitting method which was {\it different} from that employed by Hamuy et al. (1996b).  These values should not be compared directly to the values given by Hamuy et al. (1996b).  The only consistent way to combine the parameters of the SNe Ia here and in Hamuy et al. (1996b) is to fit all of these data with a single fitting method.  

    Table 4 lists information relevant to the host galaxies of the CfA
    SN Ia sample.  This includes the galaxy designation (column 2),
    the morphological type (column 3), the $B-V$ color of the host galaxy, and the offset
    between the galaxy center and the SN Ia.  The offsets were
    determined from a flux weighted centroid for the SN and the galaxy
    in $V$.  The $B-V$ galaxy colors were determined from the largest
    apertures (typically $\sim$ 20'') which avoided any foreground point source contamination.  The same size
    aperture was used to measure the galaxies magnitudes in $B$ and $V$.  The measured galaxy colors have an
    uncertainty of $\sigma=$ 0.1 mag. 

   Table 5 contains information relevant to the discovery and
   identification of each SN Ia.  This includes the discoverer and
   IAUC announcement (column 2), the method of image recording (column
   3), the date of discovery (column 4), and
   the observers who provided the spectral classification of the
   supernova (column 5).  We have obtained spectra of each of these
   SNe Ia from the FLWO and after thorough examinination we have
   found that each was of type Ia as defined by
   Branch, Fisher, \& Nugent (1993) and Filippenko (1997).  Two objects,
 SN 1995ac and SN 1995bd, displayed similar spectral characteristics as
 the peculiar SN Ia 1991T including strong Fe III and weak Si II absorption
 (Garnavich et al. 1996).

\section{Discussion}
  
  Many of these SNe Ia have been previously utilized for a variety of 
cosmological measurements as discussed in \S 1.  Nearly all of these
applications make use of estimates of the luminosity distances to
these SNe Ia.  The discussion of the most precise way to infer these
distances has evolved from the assumption of homogeneity (Leibundgut
1988; Branch \& Miller 1993; Sandage \& Tammann 1995) to methods which
account for the correlation between light-curve shape and luminosity
(Riess, Press, \& Kirshner 1995a; Hamuy et al. 1995,1996a) and employ multiple passbands to separate the effects of dust on SN Ia light
from those of luminosity (Riess, Press, \& Kirshner 1996a; Riess
et al. 1998b).  These methods are continually evolving and improving
and the specific values of the distance related parameters will likely
be superseded regularly.  For this reason, we explore characteristics
of this sample which are largely independent of the SN distances.

   In Figure 4 we show histograms of the supernova redshifts, peak
   apparent magnitude,
   epoch of first observation, number of observations, absolute
   magnitude determined from the luminosity/light-curve parameter, and line-of-sight
   extinction.  For comparison we include the same statistics for the 
27 SNe Ia from the Cal\'{a}n/Tololo Supernova Survey (C/T).
The CfA sample has a redshift range of $0.003 < z < 0.124$
   with a mean redshift of $z=0.025$. As seen in Figure 4, the
   redshifts are concentrated at lower values, though most are within
   the Hubble flow: in the rest frame of the cosmic
   microwave background (CMB), 17 of the 22 SNe Ia have $cz > 2500$ km
   s$^{-1}$. For the C/T sample the redshift range is $0.011 < z < 0.101$ with a mean of $z=0.045$.

   The following sample characteristics are derived from multi-color
   light curve shape (MLCS) fits to the $BVRI$ data as described by
   Riess, Press, \& Kirshner (1996a) and reanalyzed by Riess et al.
   (1998a); they are subject to future refinements of fitting methods.
   The fitted peak apparent magnitudes for the CfA sample
 range from  $13.16 < m_V <
   19.52$ with a mean of $m_V=15.70 \pm 1.65$.  
   For the C/T sample the range is $14.64 < m_V <
   19.35$ with a mean of $m_V=17.24 \pm 1.30$.
   The epoch of the first light curve observation for the CfA sample 
ranges from 
   12 days before maximum to 10 days after maximum.  The average
   starting epoch is coincident with $B$ maximum with half of the SNe
   beginning before this time.  The C/T sample ranges between 10 days before 
   to 12 days after maximum with a third beginning before maximum.
 Figure 4 shows a histogram of the
   absolute magnitudes as inferred from the
   MLCS light-curve shape fits on the Cepheid distance scale (Riess,
   Press, \& Kirshner 1996a). The range of luminosities implied for
   the CfA sample is
   $-19.87 < M_V < -18.80$ with a mean of $M_V=-19.40 \pm 0.28$; the C/T sample has $-19.68 < M_V < -18.81$ with a mean of $M_V=-19.27 \pm 0.29$.  
    A true SN Ia luminosity function can only be derived
   from a sample of SNe Ia with well understood selection criteria
   (Reiss et al. 1998).  
   Figure 4 also shows the distribution of visual band extinctions as
   inferred from the MLCS measurements of reddening.  This
   distribution is strongly peaked toward low extinctions, with three
   notable exceptions (SN 1995E, SN 1995bd, and SN 1996ai) each having
   more than one visual magnitude of obscuration.  One of these, SN
   1995bd, is expected to have 1.5 mag of visual extinction from the
   Milky Way Galaxy (Schlegel, Finkbeiner, \& Davis 1998).  Four more objects
   (SNe 1993ac, 1996bv, 1996bo, and 1996bk) have 0.5-1.5 mag of visual extinction.  

While the completeness and biases of the CfA SN Ia sample are not
easily defined, it is still interesting to combine these supernovae with
the C/T sample to look for patterns in a large data set. 
Figure 5 shows the extinction and light-curve shape parameters
as a function of supernova galactocentric distance and host galaxy type.
Host galaxy extinction decays rapidly with projected separation
from the nucleus and with the progression to earlier type galaxies.
The multicolor light-curve shape parameter (Riess, Press, \& Kirshner 1996a) shows that slow decline rates ($\Delta <0$)
dominate at small galactocentric distances and that there
is a general trend, first pointed out by Hamuy et al. (1996), for
faster decline rates for supernovae occurring in early-type galaxies. 
Figure 6 shows the distribution of  absolute magnitude for SNe Ia in
the Hubble flow versus projected separation from the
host galaxy.  When no correction is made for extinction or light-curve
shape, the luminosity variation is similar to that
found by Wang, H\"oflich, \& Wheeler (1997); there is a large
dispersion at small galactocentric distances
which decreases outward.
However, when the luminosity is corrected for total
extinction (Figure 6b) from MLCS, SNe
with projected separations of less than 10~kpc are found, on average,
to be brighter by about 0.3 mag than those further out. Because the projected
separation is the minimum distance the supernova can be
from the galaxy center, the few faint objects at small projected
separations could be at even larger galactocentric distances.
Elliptical hosts dominate the sample at large separations, so the
decrease of SN Ia luminosity in early-type hosts found by Hamuy et
al. (1996a) may contribute to this trend. When the
luminosity is corrected for both extinction and light-curve shape, no
trend with galactocentric distance is apparent.

   A Hubble diagram of the 17 SNe Ia from the CfA sample with $cz >
   2500$ km s$^{-1}$ is given in Figure 7.  These distances were
   derived with the MLCS method (Riess, Press, \& Kirshner 1996a)
   as prescribed by Riess et al. (1998b) in $B,V,R$, and $I$.  The
   dispersion of these distances is $\sigma=0.16$ mag.  As noted by
   Zehavi et al. (1998), the SNe Ia within $cz \approx $ 7000 km s $^{-1}$ (log$
   cz \approx 3.85$) exhibit the dynamic complement of a local void: 
   an increased expansion rate relative to the more distant SNe Ia
   (see also Tammann 1998).  
   For these SNe Ia, the difference between the expansion rates within and beyond $\sim 7000$ km s$^{-1}$ is 7\%.  It is interesting to note
   that the sense of this change in the expansion rate is opposite to
   what would be caused by a selection bias that emphasizes more luminous supernovae at larger distances.  The observed effect corresponds to a higher Hubble constant inferred locally.  If a more statistically
   significant sample of SNe Ia upholds this hint of a local void, it would help explain why SNe Ia yield
   a lower Hubble constant than distance indicators that refer to more local volumes (Freedman et al. 1998; Jacoby et al. 1992).  

   The 22 $BVRI$ SN Ia light curves presented here are composed of 1210 individual
observations, and comprise some of the most highly sampled illustrations of
the photometric history of SNe Ia in the Hubble flow.
A histogram of the number of observations for each SN Ia is shown in
Figure 4.
  Standouts include SNe 1994ae, 1995D, 1995al, 1995bd, 1995ac, and 1996X, each
with 80 to 100 observations beginning before maximum and extending 60
to 100 days after maximum.  Figure 8 shows $BVRI$ composite light
curves formed by normalizing the data in time and brightness at the
fit to the initial peak, including a correction for $1+z$ time dilation and a
$K$-correction (Hamuy et al. 1993).  By affixing the light curves at
maximum, their inhomogeneity is readily apparent.  For example, in $B$
the light curves exhibit a decline in $B$ 15 days after maximum [$\Delta m_{15}(B)$] of $0.85 < \Delta m_{15}(B) < 1.55 $ mag and in
$I$ the time and brightness of the second maximum exhibits considerable
variations.  These and other features of the light curves have been
shown to correlate with the luminosity and colors of SNe Ia (Phillips
1993; Riess, Press, \& Kirshner 1995a, 1996a; Hamuy et al. 1995,
1996a,b).  It is this property of SNe Ia which has
recently enhanced their precision as distance indicators and may lead to a
better understanding of their progenitors and physical structure
(H\"{o}flich, Wheeler \& Thielemann 1998).

We are deeply indebted to numerous observers on the 1.2 meter at Mt. Hopkins who graciously observed our supernovae and enabled us to
 construct light curves of unprecedented sampling.  We also
 wish to thank Paul Schechter and Denise Hurley who contributed to the
 compilation of these data.  The work at U.C. Berkeley was supported by
 the Miller Institute for Basic Research in Science as well as NSF grant AST-9417213, Supernova research at Harvard is supported by the NSF through grants AST-9528899 and AST-9218475.
 
\eject
\def\refitem{\par\parskip 0pt\noindent\hangindent 20pt}
\centerline {\bf References}
\vskip 12 pt

\refitem Arnett, W.D.,Branch, D., \& Wheeler, J.C. 1985, Nature 314, 337
\refitem Barbon, R., Cappellaro, E., \& Turatto, M. 1989, A\&AS, 81, 421
\refitem Beckert D.C. \& Newberry, M.V. 1989, PASP, 101, 849
\refitem Bessell, M.S. 1990, PASP, 102, 1181
\refitem Boisseau, J.R., \& Wheeler, J.C. 1991, AJ, 101, 1281
\refitem Branch, D. 1998, ARAA, in press
\refitem Branch, D., \& Miller, D. 1993, ApJ, 405, L5
\refitem Branch, D. \& Tammann, G.A. 1992, ARA\&A, 30, 359
\refitem Branch, D., Fisher, A., \& Nugent, P. 1993, AJ, 106, 2383
\refitem Cadonau, R., Sandage, A., \& Tammann, G.A. 1984, Lecture Notes in Physics, V224, 151
\refitem Cadonau, R., \& Leibundgut, B. 1990, A\&AS, 82, 145
\refitem Cappellaro, E., Turatto M., Bennetti, S., Tsvetkov, D. Yu., Bartunov, O. S., \& Makarova, I. J. 1993a, A\&A, 268, 472
\refitem Cappellaro, E., Turatto M., Bennetti, S., Tsvetkov, D. Yu,
Bartunov, O. S., \& Makarova, I. J. 1993b, A\&A, 283, 383
\refitem Cappellaro, E. et al. 1997, A\&A, 322, 431
\refitem Chincarini, G., \& Rood, H. 1977, ApJ, 214, 351
\refitem Cousins, A. W. J. 1980, S. Afr. Astron. Obs. Circ, 1, 166
 \refitem Cousins, A. W. J. 1981, S. Afr. Astron. Obs. Circ, 6, 4
\refitem  de Vaucouleurs, G. et al. 1991, in {\it  Third Reference Catalogue of Bright Galaxies} (Springer-Verlag, New York)
\refitem Della Valle, M., \& Panagia, N. 1992, AJ, 104,696
\refitem Ford, C. et al. 1993, AJ, 106, 3
\refitem Freedman, W. et al. 1998, astro-ph/9801080
\refitem Garnavich, P. M., et al. 1998a, ApJ, 493, 53
\refitem Garnavich, P. M., et al. 1998b, ApJ, in press
\refitem Garnavich, P. M., et al. 1996, AAS, 189, 4509
\refitem Goldhaber, G., et al., 1997, in {\it Thermonuclear Supernovae}, ed. P. Ruiz-Lapuente, R. Canal, \& J. Isern, Dordrecht: Kluwer, p. 777
\refitem Hamuy, M., Phillips, M. M., Suntzeff, N. B., Schommer, R. A., 
Maza, J., \& Avil\'es, R. 1996a,
AJ, 112, 2398
\refitem Hamuy, M., et al. 1996b,
AJ, 112, 2408
\refitem Hamuy, M., Phillips, M. M., Maza, J., Suntzeff, N. B., Schommer, R. A., \& Aviles, A. 1995, AJ, 109, 1
\refitem Hamuy, M., Phillips, M. M., Maza, J., Suntzeff, N. B.,
Schommer, R. A., \& Aviles, A. 1994, AJ, 108, 2226
\refitem Hamuy, M., et al. 1993, AJ, 106, 2392
\refitem Hardie, R. H., 1962, in {\it Stars and Stellar Systems, Vol. 2, Astronomical Techniques}, ed. W.A. Hiltner, (Chicago: University of Chicago Press), p. 198
\refitem Harris, W. E., Fitzgerald, M. P., \& Reed, B. C. 1981, PASP, 93,
507 
\refitem Hatano, K., Branch, D., \& Deaton, J. 1998, ApJ, 502, 177
\refitem H\"{o}flich, P., Wheeler, J. C., \& Thielemann, F. K. 1998, ApJ, 495, 617
\refitem Huchtmeier, W., \& Richter, G., 1989, in {\it A General Catalog of HI Observations of Galaxies}, (Berlin: Springer)
\refitem Jacoby, G. H., et al. 1992, PASP, 104, 599
\refitem Johnson, H. L., \& Harris, D. L., 1954, ApJ, 120, 196
\refitem Jerjen, H., \& Tammann, G.A. 1993, A\&A, 276, 1
\refitem Landolt, A.U. 1992, AJ, 104, 340
\refitem Leibundgut, B. et al. 1996, ApJ, 466, L21
\refitem Leibundgut, B. 1990, A\&A, 229, 1
\refitem Madau, P., Della Valle, M., \& Panagia, N., 1998, MNRAS, 297,17
\refitem Maza, J., Hamuy, M., Phillips, M., Suntzeff, N., \& Aviles, R. 1994, ApJ, 424, L107
\refitem Nugent, P., Phillips, M., Baron, E., Branch, D., \& Hauschildt, P., 1995, ApJ, 455, L147
\refitem Pain, R. et al., 1997, in {\it Thermonuclear Supernovae}, ed. P. Ruiz-Lapuente, R. Canal, \& J. Isern, Dordrecht: Kluwer, p. 790
\refitem Perlmutter, S., et al., 1998, Nature, 391, 51
\refitem Perlmutter, S., et al., 1997, ApJ, 483, 565
\refitem Perlmutter, S. et al., 1995, ApJ, 440, 41
\refitem Phillips, M. 1993, ApJ, L105
\refitem Riess, A. G., Nugent, P. E., Filippenko, A. V., Kirshner,
R. P., \& Perlmutter, S., 1998a, ApJ, 504, 935
\refitem Riess, A. G. et al., 1998b, AJ, 116, 1009
\refitem Riess, A. G., et al. 1997a, AJ, 114, 722
\refitem Riess, A. G., Davis, M., Baker, J., \& Kirshner, R. P. 1997b,
ApJ, 488, L1
\refitem Riess, A. G., Press W. H., \& Kirshner, R. P. 1996a, ApJ, 473, 88
\refitem Riess, A. G., Press W. H., \& Kirshner, R. P. 1996b, ApJ, 473, 588
\refitem Riess, A. G., Press W. H., \& Kirshner, R. P. 1995a, ApJ, 438, L17
\refitem Riess, A. G., Press W. H., \& Kirshner, R. P. 1995b, ApJ, 445, L91
\refitem Rust, B.W., PhD Thesis, Oak Ridge National Laboratory (ORNL-4953)
\refitem Sadakane, K. et al., 1996, PASJ, 48, 51
\refitem Saha, A. et aly, 1994, ApJ, 425, 14
\refitem Sandage, A. et al. 1996, ApJ, 460, L15
\refitem Sandage, A. et al. 1994, ApJ, 423, L13
\refitem Sandage, A., \& Tammann, G. A., 1993, ApJ, 415, 1
\refitem Sandage, A., Tammann, G. A., Panagia, N., \& Macchetto, D. 1992, ApJ, 401, L7
\refitem Schaefer, B. E., 1996, ApJ, 464, 404
\refitem Schaefer, B. E., 1995a, ApJ, 447, L13
\refitem Schaefer, B. E., 1995b, ApJ, 449, L9
\refitem Schaefer, B. E., 1994, ApJ, 426, 493
\refitem Schlegel, D. J., Finkbeiner, D. P., \& Davis, M. 1998, ApJ,
500, 525
\refitem Schmidt, B. P., et al. 1998, ApJ, in press
\refitem Shaw, R. L. 1979, A\&A, 76, 188
\refitem Smith, C. et al. 1998, in preparation
\refitem Strauss, M. A., \& Willick, J. A. 1995, PhR, 261, 271
\refitem Tammann, G. A., 1998, astro-ph/9805013
\refitem Tammann, G. A., \& Sandage, A. 1995, ApJ, 452, 16
\refitem Tammann, G. A., \& Leibundgut, B. 1990, A\&A, 236, 9
\refitem Timmes, F. X., 1991, in {\it Supernovae}, ed S.E. Woosley,
Springer Verlag, New York
\refitem Tripp, R., A\&A, 1998, 331, 815
\refitem Turatto, M., Cappellaro, E., \& Benetti, S. 1994, AJ, 108, 202
\refitem van den Bergh, S., \& McClure, R. D. 1994, ApJ, 425, 205
\refitem van den Bergh, S. 1994, ApJS, 92, 219
\refitem Vaughan, T. E., Branch, D., Miller, D.L., \& Perlmutter, S. 1995, ApJ, 439, 558
\refitem Wang, L., H\"{o}flich, P., \& Wheeler, J. C. 1997, ApJ, 483, L29
\refitem Watkins, R., \& Feldman, H. A. 1995, ApJ, 453, L73
\refitem White, M. 1998, astro-ph/9802295
\clearpage

\vfill \eject
 
{\bf Figure 1:} The solid and dotted lines show the $B,V,R,$ and $I$ CCD transmission functions for the FLWO 1.2 m telescope as determined from the FLWO passband filters and the quantum efficiency curve of the FLWO thick and thin CCD, respectively.  These are compared to the Bessell (1990) representation of the Johnson/Cousins convention for the $B,V,R,$ and $I$ phototube transmissions (dashed line).  The sharp phototube transmission functions of the Johnson/Cousins convention are not perfectly matched using CCD detectors which are sensitive to light over a range of wavelengths.

{\bf Figure 2:} Photometry comparison stars in the fields of 22 SNe
Ia.  The stars used for measuring the the brightness of each SN Ia are
listed in Table 2 and indicated in the figures.  The orientation of
each field is North at the top and East to the left.  The horizontal
arrow at the top left indicates the length of one arcminute.

{\bf Figure 3:} $B,V,R,$ and $I$ light curves of 22 SNe Ia.  The $V$
light curves (filled circles) are plotted without an offset.  The
$B$ light curves (open circles) are plotted at $\sim$ +1 mag offset from $V$, the $R$
light curves (open diamonds) are plotted at $\sim$ -1 mag offset from $V$, and the $I$
light curves (open squares) are plotted at $\sim$ -2 mag offset from
$V$.  The lines are the MLCS empirical fits to the data (Riess, Press,
\& Kirshner 1996a; Riess et al. 1998b).

{\bf Figure 4:} Characteristics of CfA SN Ia sample.  Shown with solid lines 
are
histograms of the redshifts ($z$), apparent magnitude ($m_V$), epoch of first
$V$ observation relative to $B$ maximum, number of observations,
absolute magnitude ($M_V$) as determined from the luminosity/light-curve
parameter, and line-of-sight visual extinction, $A_V$.  The
last three parameters are derived from MLCS empirical fits to the data
(Riess, Press, \& Kirshner 1996a; Riess et al. 1998b).  Shown in dash-dot lines are the same characteristics for the Cal\'{a}n/Tololo Supernova Survey (C/T).

{\bf Figure 5:} SN Ia extinction and light curve parameter trends
with galactocentric distance and host galaxy morphology.  A general
decrease of host galaxy extinction is seen with galactocentric
distance and an increase with the lateness of host galaxy type.  A
weak trend of the MLCS light-curve parameter, $\Delta$, is that
SNe Ia with comparatively faster (dimmer) light curves occur
further from the center of galaxies and are more common to early-type
galaxies.  The CfA SNe Ia are shown as filled symbols, the C/T SNe Ia
are shown as open symbols, and SNe Ia calibrated by Cepheid variables
are shown as X's.

{\bf Figure 6:} SN Ia luminosity versus galactocentric distance.  The
luminosities of SNe Ia, uncorrected for light-curve shape or
extinction (a), display a greater variation closer to the galaxy centers
as noted by Wang, H\"oflich, \& Wheeler (1997).  After correction for
extinction (b), SNe Ia with projected separations of less than 10~kpc
are, on average, brighter by about 0.3 mag than those further out.
This relation is also shown by host galaxy type (d).
After the luminosities are corrected for light-curve shape and
extinction, no significant trend with galactocentric distance is
apparent (c).  The CfA SNe Ia are shown as filled symbols, the C/T SNe Ia
are shown as open symbols. 
 
{\bf Figure 7:}  Hubble diagram of CfA sample 17 SNe Ia with $cz >
2500$ km s$^{-1}$.  The distances are determined by empirical MLCS
fits to the light curves described by Riess, Press, \& Kirshner (1996a),
and updated by Riess et al. (1998b).  

{\bf Figure 8:} Composite $B,V,R$, and $I$ SN Ia light curves.  
These light curves were made by normalizing the 22 CfA SN Ia sample
light curves in time and brightness at the
fit to the initial peak, including a correction for $1+z$ time dilation and a
$K$-correction.  They include over 1200 individual data points.  The inhomogeneity of the light curves is readily
apparent.

\vfill
\eject

\end{document}